\documentclass[aps,prb,reprint,superscriptaddress]{revtex4-1}
\usepackage{amsmath,bm,amssymb,amsfonts}
\usepackage{dcolumn}
\usepackage{graphicx}
\usepackage{latexsym}
\usepackage{epsfig}
\usepackage{multirow}

\begin{document}

\title{Topological superconductivity from transverse optical phonons in oxide heterostructures}
	
\author{Minseong Lee}
\affiliation{School of Energy and Chemical Engineering, Ulsan National Institute of Science and Technology, Ulsan, 44919, Korea}
\author{Hyun-Jae Lee}
\affiliation{School of Energy and Chemical Engineering, Ulsan National Institute of Science and Technology, Ulsan, 44919, Korea}
\author{Jun Hee Lee}
\email{junhee@unist.ac.kr}
\affiliation{School of Energy and Chemical Engineering, Ulsan National Institute of Science and Technology, Ulsan, 44919, Korea}
\author{Suk Bum Chung}
\email{sbchung0@uos.ac.kr}
\affiliation{Department of Physics, University of Seoul, Seoul 02504, Korea}
\affiliation{Natural Science Research Institute, University of Seoul, Seoul 02504, Korea}
\affiliation{School of Physics, Korea Institute for Advanced Study, Seoul 02455, Korea}
\begin{abstract}
A topological superconductor features at its boundaries and vortices Majorana fermions, which are potentially applicable for topological quantum computations. The scarcity of the 
known experimentally verified physical systems with topological superconductivity, time-reversal invariant ones in particular, is giving rise to a strong demand for 
identifying new candidate materials. In this research, we study a heterostructure consisting of a transition metal oxide two-dimensional electron gas (2DEG) 
sandwiched by 
insulators near the paraelectric (PE) / ferroelectric (FE) phase transition. Its relevant characteristics is the combination of 
the transition metal spin-orbit coupling and the 
soft odd-parity phonons 
arising from the ferroelectric fluctuation; 
it gives rise to the 
fluctuating Rashba 
effect, which 
can mediate the pairing interaction for time-reversal invariant topological superconductivity. 
As 
the PE / FE phase transition can be driven by applying strain on the heterostructure, this system provides a tunable electron-phonon coupling. 
Through the first-principle calculations on the (001) [BaOsO$_3$][BaTiO$_3$]$_4$, 
we find such electron-phonon coupling to be strong over a wide range of applied tensile bi-axial strain in the monolayer BaOsO$_3$ sandwiched between the (001) BaTiO$_3$, 
hence qualifying it as a good candidate material. 
Furthermore, the stability of topological superconductivity in this material is enhanced by its orbital physics that gives rise to the anisotropic dispersion. 
\end{abstract}
\maketitle

\section{Introduction}

In recent years, much of the research on superconductivity has been focusing on its topological properties \cite{Hasan2010rev, Qi2011rev}. 
Historically, such approach began with recognizing the topological nature of the $^3$He superfluidity, including the time-reversal invariant phases \cite{Salomaa1988, Volovik2009bk}. The subsequent development of topological insulator led to the current theoretical understanding that all superconducting states can be classified according to their topology \cite{Read2000, Qi2009, Schnyder2008, Roy2006}. 
The chief characteristics of the topologically non-trivial (henceforth `topological') superconductor is the existence of Majorana fermions at its boundaries and the cores of various defects \cite{Read2000, Wilczek2009}; their potential application to topological quantum computation has enormously stimulated the interest in topological superconductivity (TSC) \cite{Nayak2008rev, Alicea2012rev}.

One challenge in this field has been obtaining material candidates for time-reversal invariant (TRI) TSC. While 
various proposals to achieve TSC `extrinsically', {\it i.e.} from proximity to a topologically trivial external superconductor \cite{Fu2008, Sato2009, Sau2010, Alicea2010} 
have led to experimental realizations, {\it e.g.} 
\cite{Mourik2012, Nadj-Perge2014}, 
they involve breaking time-reversal symmetry and hence do not lead to 
TRI TSC. Therefore an important challenge in this field now is to devise a proposal for `intrinsic' TRI TSC, 
{\it i.e.} finding a candidate material 
where the pairing interaction and the electronic band structure for TRI TSC exist.

Recently, it has been shown that the inversion symmetry breaking fluctuation can mediate the pairing interaction for the TRI TSC \cite{Kozii2015, YWang2016}. In terms of pairing symmetry, the intrinsic TSC often requires the odd parity 
rather than the even parity as in the conventional ($s$-wave) and the cuprate ($d$-wave) superconductors. 
A class of odd parity phonons can be classified as such inversion symmetry breaking fluctuations. 
As an example, 
the continuous inversion symmetry breaking transition from the paraelectric (PE) phase to the ferroelectric (FE) phase 
arises from the continuous change in the crystalline structure, which requires softening of 
the transverse optical phonon modes 
near the transition point; 
there would be a strong phonon mediated electronic interaction in this region \cite{Chandra2017}. While it has been shown that the pairing interaction arising from the electron-phonon coupling alone is always strongest in the $s$-wave channel \cite{Scheurer2016}, 
it has been shown that for the isotropic dispersion, the $s$-wave superconductivity can be energetically degenerate with the $p$-wave TRI TSC \cite{Kozii2015, YWang2016}. In such case, 
additional 
perturbations can easily 
stabilize the 
$p$-wave over the $s$-wave pairs, leading to 
the TRI TSC in vicinity of 
the PE / FE transition. 
However, materials known to be superconducting in vicinity of the PE / FE transition either have a complicated, and therefore anisotropic, band structure \cite{Sakai2001, Matsubayashi2018} or a low density of state arising only out of doping \cite{VanDerMarel2011, Edge2015}, which would suppress the superconducting $T_c$ in absence of fine-tuning. 

In this paper, we show that the isotropic band dispersion 
is not necessary 
for topological superconductivity driven by odd-parity phonons and, from the first principle calculation results, present as a candidate material 
the single layer of BaOsO$_3$ sandwiched between the (001) BaTiO$_3$, where 
a high-density 2DEG arising in the BaOsO$_3$ monolayer has been shown to features bands with strongly anisotropic, {\it i.e.} quasi-one-dimensional (1d),  dispersions \cite{ZZhong2015}. We will explain that 
it is thanks to these quasi-1d bands that the TRI TSC may be realized in this material; while they have been shown to have a tendency toward the $p$-wave superconductivity with the Hubbard interaction \cite{Cho2013}, we are finding a similar result for the interaction mediated by the TO$_1$ phonons. A key feature in our candidate material is the PE / FE transition that can be induced by applying bi-axial tensile strain on the $ab$-plane, where the $c$-axis polarized TO$_1$ phonons would soften \cite{Dieguez2005, Schlom2007, JHLee2008} and mediate strong electronic interaction. As the BaTiO$_3$ ferroelectricity has been shown to induce a giant switchable Rashba effect on the osmate 2DEG \cite{ZZhong2015}, we can expect the $c$-axis polarized TO$_1$ phonons to couple strongly to the 2DEG over a large range of tensile strain. 
In possessing a tuning parameter for the electronic interaction, our material is analogous to the twisted bilayer graphene \cite{YCao2018}. The close analogy between the band dispersion of 
the osmate 2DEG and Sr$_2$RuO$_4$ 
leads us to consider the simplified model for the Sr$_2$RuO$_4$ bands and consider its coupling 
to the $c$-axis polarized TO$_1$ phonons along with the possibility of the TRI TSC arising from such electron-phonon coupling. We will then discuss our first-principle calculation results for both the lattice and the band structure of the (001) [BaOsO$_3$][BaTiO$_3$]$_4$ superlattice, the former showing the tensile strain induced transition in the $c$-axis electric polarization and the latter including the coupling to the $c$-axis TO$_1$ phonons. Our qualitative comparison between the first-principle calculation on the (001) [BaOsO$_3$][BaTiO$_3$]$_4$ superlattice and our simplified model for the Sr$_2$RuO$_4$ band indicates that the monolayer BaOsO$_3$ sandwiched between the (001) BaTiO$_3$ would be a good candidate material for realizing the TRI TSC mediated by the TO$_1$ phonons.

\section{The TO$_1$ phonon induced superconductivity in the simplified ruthenate model}

We first consider 
the band structure of Sr$_2$RuO$_4$, which we shall treat as purely two-dimensional (2d). To derive the coupling between the electrons and the $c$-axis polarized TO$_1$ phonons, we will first consider how this band structure would be modified by the inversion symmetry breaking, specifically by displacing the Ru atom away from the center of the O octahedron along the $c$-axis. 
We will then discuss the competition between the topologically trivial superconductivity and the TRI TSC arising from this electron-phonon coupling. 

\subsection{The electronic structure and its coupling to the TO$_1$ phonons}

\begin{figure}
        \includegraphics[width=1.0\columnwidth]{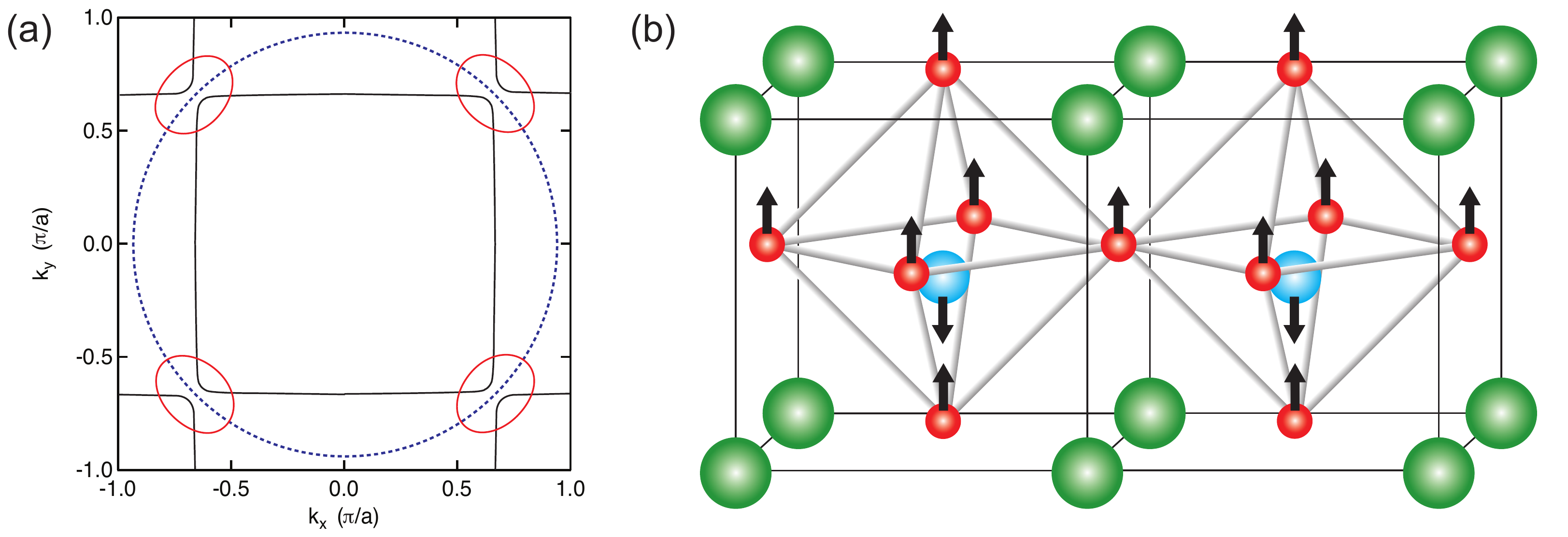}
	\caption{(a) The schematic Fermi surfaces of Sr$_2$RuO$_4$; note two pockets with nearly 1d dispersion save for the avoided band crossing regions marked in red \cite{Chung2012}. (b) The relative motion of atoms (the blue, green and red represents Ru, Sr, and O, respectively) in the TO$_1$ phonon mode breaks the inversion symmetry with respect to the $ab$-plane.}
	\label{FIG:band}
\end{figure}

The two characteristics of the Sr$_2$RuO$_4$ electronic band structure that are relevant to the TRI TSC are shown in Fig.~\ref{FIG:band} (a); one is the existence of three pockets at the Fermi level and the other the quasi-1d dispersions in two of these pockets. 
Possessing an odd number of pockets with spin degeneracy at the Fermi level is one necessary condition for for the TRI TSC 
\cite{Qi2010} \footnote{One should halve the number of Fermi surface when inversion symmetry breaking removes the spin degeneracy.}; with an even number of Fermi surfaces, non-trivial topology is possible only with the additional symmetry protection 
({\it e.g.} the crystalline symmetry \cite{Fu2011, Teo2013})  
regardless of the pairing symmetry. These three pockets all arises from the three Ru $t_{2g}$ ($d_{xy}, d_{xz}, d_{yz}$) orbitals 
\cite{Mackenzie2003rev, Haverkort2008}. Fig.~\ref{FIG:band} (a) allows us to divide these pockets into two group - the square-like (rounded only at edges) $\alpha, \beta$ around $\Gamma$, M respectively and the more circular $\gamma$ around $\Gamma$; 
this suggests that $\alpha$ and $\beta$ 
arise from the hybridization two quasi-1d bands 
\cite{Raghu2010, Chung2012}. 
The dispersion shown in Fig.~\ref{FIG:band} (a) 
can be 
explained by the symmetries of the $t_{2g}$ orbitals on the square lattice; they 
suppress the nearest-neighbor hopping intra-orbital of the $d_{xz}$ ($d_{yz}$) orbital in the $y(x)$-direction, 
resulting in the quasi-1d dispersion for these orbitals and forbid the inter-orbital nearest-neighbor hopping between different $t_{2g}$ orbitals, 
reducing the orbital hybridization. 
Likewise, they also ensure that the nearest-neighbor hopping between the $d_{xy}$ orbital is equally strong in both the $x$- and $y$-direction, 
hence the more isotropic dispersion of the $\gamma$ pocket. For our purpose, the relevant hybridizations between the $t_{2g}$ orbitals 
arises from the onsite atomic spin-orbit coupling. 
So the Hamiltonian in the $t_{2g}$ orbital basis $\{|d_{xz}\rangle, |d_{yz}\rangle, |d_{xy}\rangle\}$ can be written as
\begin{equation}
\mathcal{H}_{\rm hop} ({\bf k}) + \eta \mathcal{H}_{\rm SOC} \!=\! \left(\begin{array}{ccc}\xi_{xz}({\bf k}) & -i\sigma_z\eta & i\sigma_x\eta\\
                                                             i\sigma_z\eta & \xi_{yz}({\bf k}) & -i\sigma_y\eta\\
                                                             -i\sigma_x\eta & i\sigma_y\eta & \xi_{xy}({\bf k}) + \delta\end{array}\right),
\label{EQ:invSymm}
\end{equation}
where $\sigma_{x,y,z}$ are the spin Pauli matrices, $\xi_{xz/yz} = -2t \cos k_{x/y} - 2t^\perp \cos k_{y/x}$, and $\xi_{xy} = -2t (\cos k_x + \cos k_y) - 4t'\cos k_x \cos k_y$. 
In addition to the small $t^\perp/t, t'/t$ attributable to the $t_{2g}$ orbital symmetries, 
the atomic spin-orbit coupling Ru gives us a similarly small $\eta/t$ in Sr$_2$RuO$_4$ \cite{Haverkort2008}. 
Hence, 
all bands can be considered to be in the orbital eigenstates except for the comparatively small region in vicinity of the avoided band crossing marked in Fig.~\ref{FIG:band} (a). 
So far, the inversion symmetry with respect to the $ab$-plane has been assumed, 
with each Ru atom at the center of the O octahedron.


Our aims is to examine the superconductivity arising from the coupling of the above band structure to the $c$-axis polarized TO$_1$ phonon mode shown in Fig.~\ref{FIG:band} (b), which we will derive by considering how the band structure is modified by the inversion symmetry breaking due to the uniform displacement of the Ru atom away from the center of the O octahedron along the $c$-axis. The most immediate effect $\hat{\mathcal{H}}_\mathrm{ISB}$ of in this case consists of the spin-conserving inter-orbital nearest neighbor hopping between the orbitals odd in the inversion with respect to the $ab$-plane ($d_{xz}$, $d_{yz}$) and even in this inversion ($d_{xy}$, $d_{z^2}$, $d_{x^2-y^2}$) \cite{Shanavas2014l}. While the effect of the inversion symmetry breaking is entirely {\it local} in this approach, the effect of the uniform inversion symmetry breaking can be described using only 3 parameters 
\begin{eqnarray}
\gamma_1 \equiv& \langle d_{xy} | \hat{\mathcal{H}}_\mathrm{ISB} | d_{xz} \rangle_{\bf \hat{y}} = \langle d_{xy} | \hat{\mathcal{H}}_\mathrm{ISB} | d_{yz} \rangle_{\bf \hat{x}},\nonumber\\
\gamma_2 \equiv& \langle d_{xz} | \hat{\mathcal{H}}_\mathrm{ISB} | d_{z^2} \rangle_{\bf \hat{x}} = \langle d_{yz} | \hat{\mathcal{H}}_\mathrm{ISB} | d_{z^2} \rangle_{\bf \hat{y}},\nonumber\\
\gamma_3 \equiv& \langle d_{xz} | \hat{\mathcal{H}}_\mathrm{ISB} | d_{x^2-y^2} \rangle_{\bf \hat{y}} = \langle d_{yz} | \hat{\mathcal{H}}_\mathrm{ISB} | d_{x^2-y^2} \rangle_{\bf \hat{x}},
\label{EQ:ISB}
\end{eqnarray}
in which the vectors in the subscripts denote the relative position of the two orbitals. 
All these inversion symmetry breaking hoppings are 
odd in parity, {\it e.g.} $\langle d_{xy} | \hat{\mathcal{H}}_\mathrm{ISB} | d_{xz} \rangle_{-{\bf \hat{y}}}=-\langle d_{xy} | \hat{\mathcal{H}}_\mathrm{ISB} | d_{xz} \rangle_{\bf \hat{y}}$, and can be considered as the generalizations of the chiral orbital angular momentum coefficients \cite{Park2011}.  Recent first-principle calculations find $|\gamma_{2,3}| \gg |\gamma_1|$ for the transition metal oxide $t_{2g}$ electrons with strong Rashba effect \cite{Shanavas2014l, Kim2016} due to the $c$-axis component of the local electric field generated by the relative displacement between the transition metal atom and the oxygen atom \cite{Shanavas2014b, Shanavas2016}; hence, the necessity of including the $e_g$ orbitals ($d_{z^2}$ and $d_{x^2-y^2}$) away from the Fermi surfaces in analyzing effects of the inversion symmetry breaking. 

On the Fermi surface, the lowest order effect of $\hat{\mathcal{H}}_{\rm ISB}$ in presence of $\eta \hat{\mathcal{H}}_{\rm SOC}$ manifests itself as the 
Rashba effect which breaks the spin degeneracy and is odd in parity yet preserves time-reversal symmetry. From the perturbation theory, the lowest order effect of $\hat{\mathcal{H}}_{\rm ISB}$, which is ${\bf k}$-odd, on each orbital 
gives rise to the effective Rashba Hamiltonian 
\begin{equation}
\mathcal{H}^{i}_{\rm R} ({\bf k}) \approx \lambda_i {\bm \sigma}\cdot {\bf d}_{\bf k}^{i},
\label{EQ:hamR_eff}
\end{equation}
where ${\bf d}_{\bf k}$ is a dimensionless vector odd in ${\bf k}$ and perpendicular to ${\bf \hat{z}}$. Due to the weak inversion symmetry breaking effects on the $d_{xy}$ orbital, the strength of the Rashba strength on the $d_{xz} / d_{yz}$ orbitals, $\lambda_{xz} = \lambda_{yz} =4\eta( -\sqrt{3}\gamma_2/\delta\epsilon_{z^2/xz} + \gamma_3/\delta\epsilon_{x^2-y^2/xz})$, $\delta\epsilon$'s being the energy difference between the orbitals at the nearest time-reversal invariant momentum, dominates over that of the $d_{xy}$ orbital. We further note here that ${\bf d}_{\bf k}^{xz} = {\bf \hat{y}}\sin k_x$, ${\bf d}_{\bf k}^{yz} = -{\bf \hat{x}}\sin k_y$ give us a quasi-1d Rashba effects on the $d_{xz} / d_{yz}$ orbitals with essentially uniform spin states and magnitude over the Fermi surface. 

The above band structure modification can be used to obtain the coupling to the TO$_1$ phonons, as it is equivalent to 
the $q=0$ $c$-axis polarized TO$_1$ phonon mode 
of Fig.~\ref{FIG:band} (b). 
Therefore, this electron-phonon coupling should 
be given by the generalization of
the inversion symmetry breaking inter-orbital hoppings considered in Eq.~\eqref{EQ:ISB}. To obtain this generalization, we first observe that all the parameters of Eq.~\eqref{EQ:ISB}, as they arise from the inversion symmetry breaking, would have their signs reversed if we reverse the sign of the Ru-O relative displacement; hence it would be reasonable to take these parameters to be proportional to the TO$_1$ phonon displacement and not the displacement gradient as in the coupling to acoustic phonons. Furthermore, given that the effect of the Ru-O relative displacement is local, {\it i.e.} inducing the nearest neighbor hopping, ignoring any effects of the spatial variation of the phonon displacement would be a reasonable approximation. This substitution of Eq.~\eqref{EQ:ISB} by spatially varying {\it local} hopping parameters, {\it i.e.} $\gamma_{1,2,3} \to \tilde{\gamma}_{1,2,3} Q_{\bf r}$ where $Q_{\bf r}$ is the local TO$_1$ phonon mode displacement at the position ${\bf r}$, leads to the effective electronic coupling to the TO$_1$ phonons at each orbital to be of the Yukawa-type \cite{YWang2016}  
\begin{equation}
\hat{\mathcal{H}}_{\rm e\!-\!ph}^i \!=\!  \frac{1}{2\sqrt{N}}\tilde{\lambda}_i\sum_{{\bf k},{\bf q}}\sum_{s,s'}  Q_{\bf q} c^\dagger_{{\bf k}+{\bf q},i,s}[{\bm \sigma}\cdot({\bf d}^i_{\bf k}+{\bf d}^i_{{\bf k}+{\bf q}})]_{s,s'} c_{{\bf k},i,s'},
\label{EQ:TO1Couple}
\end{equation} 
where $i$ is the orbital index, $\tilde{\lambda}_{xz} = \tilde{\lambda}_{yz} = 4\eta(-\sqrt{3}\tilde{\gamma}_2/\delta\epsilon_{z^2/xz} + \tilde{\gamma}_3/\delta\epsilon_{x^2-y^2/xz})$, $N$ the total number of unit cells on the $ab$-plane, and $c$ ($c^\dagger$) the electron annihilation (creation) operator. Given our derivation of the electron-phonon coupling, the $d_{xy}$ band with its weak Rashba effect would also have a weak electron-phonon coupling. However, our Rashba-type electron-phonon coupling as a whole preserves the inversion symmetry, both $Q_{\bf q}$ and ${\bf d}_{\bf k}$ being odd under inversion. Hence this electron-phonon coupling exists 
regardless of the inversion symmetry breaking; for the case of the broken inversion symmetry, the Rashba effect of Eq.~\eqref{EQ:hamR_eff} is added to the band structure. 
As we are treating our band structure to be purely 2d, it would be consistent 
to ignore any charge screening effect of $c$-axis polarized phonon modes.

Eq.~\eqref{EQ:TO1Couple} implies that the softer the TO$_1$ phonons, the stronger their coupling to the electrons. When the harmonic approximation is applicable, the phonon mode annihilation and creation operator $a$ and $a^\dagger$ 
are related to the mode displacement $Q$ by $a_{\bf q} + a^\dagger_{-{\bf q}}=\sqrt{2M_{\rm TO_1}\omega_{\bf q}/\hbar}Q_{\bf q}$, with $M$ and $\omega_{\bf q}$ being the effective phonon mass and  the mode frequency, respectively. Therefore, in terms of $a$ and $a^\dagger$, we obtain the electron-phonon coupling on each orbital of   
\begin{align}
\hat{\mathcal{H}}_\mathrm{e-ph}^i \approx \frac{1}{\sqrt{N}}  \sum_{{\bf k},{\bf q}}\sum_{s,s'}& g^i_{\bf q} (a_{\bf q} + a^\dagger_{-{\bf q}})\nonumber\\
&\times c^\dagger_{{\bf k}+{\bf q},i,s} [{\bm \sigma}\cdot({\bf d}^i_{\bf k}+{\bf d}^i_{{\bf k}+{\bf q}})]_{s,s'} c_{{\bf k},i,s'},
\label{EQ:momCouple31}
\end{align}
where 
$$
g^i_{\bf q} = \left\{\begin{array}{cc} \frac{1}{2}\tilde{\lambda}_{xz}\sqrt{\frac{\hbar}{2M_{\rm TO_1}\omega_{\bf q}}}, & i=xz,yz\\
0, & i=xy \end{array}\right.
$$
is the effective electron-phonon coupling parameter to the TO$_1$ phonon mode at ${\bf q}$. We see here that the magnitude of the electron-phonon coupling is exactly that of the Rashba coefficient for half the quantum root-mean square zero-point displacement of the TO$_1$ mode, and therefore grows with the zero-point displacement at the lower phonon frequency. The TO$_1$ phonon softening therefore 
leads to its electronic coupling 
dominating the electron correlation. As the above derivation works best at the long-frequency limit, we shall henceforth assume that the phonon frequency minimum occurs at $q=0$. 

\subsection{TO$_1$ phonon induced quasi-1D superconductivity}

The superconductivity induced by the coupling to the TO$_1$ phonons is mostly determined by 
the electron-electron interaction mediated by the TO$_1$ phonons on the Fermi surface in the Cooper channel. 
We obtain \cite{YWang2016} 
\begin{align}
\mathcal{U}^{(Cooper)}_{\rm e-e} \!=\!& -\frac{1}{N}\sum_{{\bf k},{\bf k}'} \sum_{s,s',\tau,\tau'} V_0({\bf k}-{\bf k}')  c^\dagger_{{\bf k}',s} c^\dagger_{-{\bf k}',\tau} c_{-{\bf k},\tau'} c_{{\bf k},s'}\nonumber\\
&\times \left(\frac{{\bf d}_{\bf k} + {\bf d}_{{\bf k}'}}{2}\cdot{\bm \sigma}\right)_{s,s'} \left(\frac{{\bf d}_{-{\bf k}} + {\bf d}_{-{\bf k}'}}{2}\cdot{\bm \sigma}\right)_{\tau,\tau'}.
\label{EQ:pairInt1}
\end{align}
where $V_0 ({\bf q})=8|g_{\bf q}|^2/\hbar \omega_{\bf q}=\tilde{\lambda}^2/M_{\rm TO_1}\omega_{\bf q}^2 > 0$, by applying the second-order perturbation theory on the electron-phonon coupling of Eq.~\eqref{EQ:momCouple31}.  Given that the low momentum Cooper pair scattering will be dominant with the frequency minimum occurring at at $q=0$, our analysis will be confined to the intra-orbital interaction and the orbital indices will be omitted. 
The stronger electron-phonon coupling of the quasi-1d bands means that the superconductivity mediated by the TO$_1$ phonon would be quasi-1d, yet without, however, the spontaneous time-reversal symmetry breaking of the 
previous model for Sr$_2$RuO$_4$ \cite{Raghu2010}.

When the inversion symmetry is preserved, the above interaction of Eq.~\eqref{EQ:pairInt1} gives rise to the pairing interactions in both the $s$-wave spin-singlet channel and the odd parity spin-triplet channel parametrized by ${\bf d}_{\bf k}$. This can be shown most simply in the helicity basis, defined as
\begin{equation}
{\bf d}_{\bf k} \cdot {\bm \sigma} c_{{\bf k},\pm} = \pm |{\bf d}_{\bf k}| c_{{\bf k},\pm}.
\label{EQ:helicity0}
\end{equation}
In this helicity basis, the $s$-wave spin-singlet pairing and the odd parity spin-triplet pairing parameterized by ${\bf d}_{\bf k}$ are both intra-helicity pairings \cite{Frigeri2004, YWang2016, Smidman2017},
\begin{align}
\sum_{s,s'} (i\sigma^y)_{s,s'} c^\dagger_{{\bf k},s} c^\dagger_{-{\bf k},s'} \!=&\! -\!\sum_{\nu=\pm} e^{-i\nu\phi_{\bf k}} c^\dagger_{{\bf k},\nu} c^\dagger_{-{\bf k},\nu},\nonumber\\
\sum_{s,s'} [({\bf \hat{d}}_{\bf k} \cdot {\bm \sigma})i\sigma^y]_{s,s'}c^\dagger_{{\bf k},s} c^\dagger_{-{\bf k},s'} \!=&\! -\!\sum_{\nu=\pm} \nu e^{-i\nu\phi_{\bf k}}c^\dagger_{{\bf k},\nu} c^\dagger_{-{\bf k},\nu}
\label{EQ:singletTriplet}
\end{align} 
respectively, where ${\bf \hat{d}}_{\bf k} \equiv {\bf d}_{\bf k}/|{\bf d}_{\bf k}|$ and  $e^{i\phi_{\bf k}} \equiv {\bf \hat{d}}_{\bf k} \cdot ({\bf \hat{x}}+i{\bf \hat{y}})$. Eq.~\eqref{EQ:singletTriplet} identifies the relative sign between the two intra-helicity pairing gaps as the difference between the $s$-wave and the odd-parity pairing in the helicity basis. Two competing superconducting states can be identified with this relative sign of the two pairing gaps even when the inversion symmetry breaking allows for the magnitude of the two intra-helicity pairing gaps to differ, which hybridizes the $s$-wave spin-triplet pairing and the odd parity spin-triplet pairing \cite{Qi2010, Smidman2017}; henceforth the `$s$-wave' and the `odd parity' will be understood in the sense of this relative sign. From the transformation from the spin basis to the helicity basis,
\begin{align}
&c^\dagger_{{\bf k}'} \frac{{\bf d}_{\bf k} + {\bf d}_{{\bf k}'}}{2}\cdot{\bm \sigma}  c_{\bf k}\nonumber\\ =& \frac{|{\bf d}_{\bf k}|+|{\bf d}_{{\bf k}'}|}{2} \sum_{\nu=\pm}\frac{1+e^{-i\nu(\phi_{{\bf k}'}-\phi_{\bf k})}}{2} c^\dagger_{{\bf k}',\nu} c_{{\bf k},\nu}\nonumber\\
+&  \frac{|{\bf d}_{\bf k}|-|{\bf d}_{{\bf k}'}|}{2}\sum_{\nu=\pm}\frac{-e^{-i\nu\phi_{\bf k}}+e^{-i\nu\phi_{{\bf k}'}}}{2} c^\dagger_{{\bf k}',\nu} c_{{\bf k},-\nu},
\label{EQ:helicity1}
\end{align}
we can see that 
the inter-helicity scattering term vanishes if either $|{\bf d}_{\bf k}|$ or ${\bf \hat{d}}_{\bf k}$ is constant over the Fermi surface. In such cases, 
\begin{align}
\mathcal{U}^{(Cooper)}_{\rm e-e} \!=&\!-\!\frac{1}{N}\sum_{{\bf k},{\bf k}'} V_0 ({\bf k}'-{\bf k}) \frac{1+{\bf \hat{d}}_{\bf k} \cdot {\bf \hat{d}}_{{\bf k}'}}{2}\left(\frac{|{\bf d}_{\bf k}|+|{\bf d}_{{\bf k}'}|}{2}\right)^2\nonumber\\ 
&\times \sum_{\nu=\pm} e^{-i\nu(\phi_{{\bf k}'}-\phi_{\bf k})}c^\dagger_{{\bf k}',\nu} c^\dagger_{-{\bf k}',\nu} c_{-{\bf k},\nu} c_{{\bf k},\nu},
\end{align}
shows that the two intra-helicity pairings, the $s$-wave and the odd parity, are exactly degenerate. 

Stabilizing the odd parity pairing over the $s$-wave pairing requires additional perturbation \cite{YWang2016, Scheurer2016}, 
the 
repulsive Hubbard interaction $\mathcal{U}_H = U \sum_{\bf r} n_{{\bf r}\uparrow} n_{{\bf r}\downarrow}$ being the simplest example that preserves time-reversal symmetry. In absence of such interaction, when $|{\bf d}_{\bf k}|$ or ${\bf \hat{d}}_{\bf k}$ is not exactly constant over the entire Fermi surface, the $s$-wave / odd parity degeneracy is broken by the inter-helicity pair tunneling arising from $\mathcal{H}^{(Cooper)}_{\rm e-e}$
\begin{align}
\mathcal{U}^{(tunnel)}_{\rm e-e} \!=&\! -\!\frac{1}{N}\sum_{{\bf k},{\bf k}'} V_0 ({\bf k}'-{\bf k}) \frac{1-{\bf \hat{d}}_{\bf k} \cdot {\bf \hat{d}}_{{\bf k}'}}{2}\left(\frac{|{\bf d}_{\bf k}|-|{\bf d}_{{\bf k}'}|}{2}\right)^2\nonumber\\
&\times\sum_{\nu=\pm} e^{-i\nu(\phi_{{\bf k}'}+\phi_{\bf k})}c^\dagger_{{\bf k}',\nu} c^\dagger_{-{\bf k}',\nu} c_{-{\bf k},-\nu} c_{{\bf k},-\nu},
\label{EQ:pairTunnel}
\end{align}
which favors the $s$-wave pairing or, more precisely, the positive relative sign between the intra-helicity pairings of Eq.~\eqref{EQ:singletTriplet}. By contrast, from the helicity basis representation of the 
Hubbard interaction in the Cooper channel, 
\begin{equation}
\mathcal{U}^{(Cooper)}_H 
\!=\! \frac{U}{4N}\!\sum_{{\bf k}, {\bf k}'} \sum_{\nu,\nu'=\pm}\!e^{-i(\nu\phi_{{\bf k}'}-\nu'\phi_{\bf k})} c^\dagger_{{\bf k}',\nu} c^\dagger_{-{\bf k}',\nu}   c_{-{\bf k},\nu'} c_{{\bf k},\nu'},
\end{equation}
one can see that 
condition reversing the effect of 
the inter-helicity tunneling $\mathcal{U}^{(tunnel)}_{\rm e-e}$ of Eq.~\eqref{EQ:pairTunnel} without canceling out the pairing interaction $\mathcal{U}^{(Cooper)}_{\rm e-e}$, hence stabilizing the odd parity pairing over the $s$-wave pairing, would be
\begin{widetext}
\begin{equation}
\frac{\tilde{\lambda}^2}{2M_{\rm TO_1}}\langle \omega_{{\bf k}-{\bf k}'}^{-2}(1-{\bf \hat{d}}_{\bf k} \cdot {\bf \hat{d}}_{{\bf k}'})(|{\bf d}_{\bf k}|-|{\bf d}_{{\bf k}'}|)^2\rangle_{\rm F.S.} <U < \frac{\tilde{\lambda}^2}{2M_{\rm TO_1}}\langle \omega_{{\bf k}-{\bf k}'}^{-2} (1+{\bf \hat{d}}_{\bf k} \cdot {\bf \hat{d}}_{{\bf k}'})(|{\bf d}_{\bf k}|+|{\bf d}_{{\bf k}'}|)^2\rangle_{\rm F.S.},
\label{EQ:oddStable}
\end{equation} 
\end{widetext}
where $\langle \cdots \rangle_{\rm F.S.}$ is the Fermi surface average over both ${\bf k}$ and ${\bf k}'$. Given that $\omega_{\bf q}^{-2}$ diverges at $q=0$ as the TO$_1$ phonons soften, when either $|{\bf d}_{\bf k}|$ or ${\bf \hat{d}}_{\bf k}$ is nearly constant over the Fermi surface, the odd parity pairing would be stabilized over the $s$-wave pairing for a wide range of $U$. 

Thanks to the nearly constant $\hat{\bf d}_{\bf k}$ over its Fermi surfaces, 
a model for the 
TRI TSC can be obtained from the $d_{xz} / d_{yz}$ orbital quasi-1D bands of Sr$_2$RuO$_4$ coupled to the TO$_1$ phonons; 
previously such possibility 
has been discussed only for the constant $|{\bf d}_{\bf k}|$ arising from the isotropic dispersion \cite{Kozii2015, YWang2016}. In deriving Eq.~\eqref{EQ:hamR_eff}, we found $\hat{\bf d}_{\bf k} = {\bf \hat{y}}$ ($\hat{\bf d}_{\bf k} = -{\bf \hat{x}}$) for the $d_{xz}$ ($d_{yz}$) orbital Fermi surface, and hence, from Eq.~\eqref{EQ:momCouple31}, the inter-helicity scattering induced by the TO$_1$ phonons is strongly suppressed, which, from Eq.~\eqref{EQ:oddStable}, allows for a repulsive Hubbard interaction to stabilize the odd parity pairing of Eq.~\eqref{EQ:singletTriplet} over a wide range of magnitude. 
While the odd parity pairing directly leads to the TRI TSC in the single-band isotropic model, the multi-band effect should be considered for the topology of our quasi-1d model. 
At its zeroth order, 
we can see from Eq.~\eqref{EQ:singletTriplet} that, while both $d_{xz}$ and $d_{yz}$ Fermi surfaces have the equal superpositions of $|\!\uparrow\uparrow\rangle$ and $|\!\downarrow\downarrow\rangle$, their relative phase factors between 
$|\!\uparrow\uparrow\rangle$ and $|\!\downarrow\downarrow\rangle$ 
differ by -1.  
While the orbital component of this TO$_1$ phonon driven quasi-1d model is closely analogous to the quasi-1d model for the chiral superconductivity of Sr$_2$RuO$_4$ \cite{Raghu2010}, 
this orbital dependence of the Cooper pair spin state requires that 
in order for the Fermi surfaces to be fully gapped for both spin-up and spin-down electrons  
both the $|\!\uparrow\uparrow\rangle$ and the $|\!\downarrow\downarrow\rangle$ pairings should be chiral but with the chirality {\it opposite} to each other, 
thus preserving the time-reversal symmetry. 
The two quasi-1D orbitals 
by themselves would give us a TRI topological crystalline superconductivity 
with the non-trivial topology protected by both the time-reversal symmetry and 
the crystalline symmetry \cite{Fu2011, Teo2013}; 
the inclusion of the $d_{xy}$ orbital is necessary for the TRI TSC 
regardless of the crystalline symmetry \cite{Qi2010}.

The stability of the TRI TSC in the quasi-1d model, compared to that of the isotropic model, is less affected when the inversion symmetry breaking removes the spin degeneracy at the Fermi surface through the Rashba effect. 
This is due to the different effects that the Fermi surface splitting have on these two models. 
It was pointed out this turns on the inter-helicity electron scattering for the isotropic model at the Fermi surface, 
as $|{\bf d}_{\bf k}|$'s always differ between the Fermi surfaces for the $\pm$ helicities in Eq.~\eqref{EQ:helicity1}, thus enhancing the $s$-wave pairing \cite{YWang2016}. 
By contrast, 
the 
presence of the Rashba term will not affect the uniformity of ${\bf \hat{d}}_{\bf k}$ from ${\bf d}^{xz}_{\bf k} = {\bf \hat{y}}\sin k_x$ (${\bf d}^{yz}_{\bf k} = -{\bf \hat{x}}\sin k_y$) in the quasi-1d model, except through enlarging the region around the avoided band crossing. Therefore, by Eq.~\eqref{EQ:helicity1}, 
the enhancement of the intra-orbital inter-helicity scattering 
due to the inversion symmetry breaking will be weak for much of the Fermi surfaces. We note that our discussion so far require material characteristics such as 2DEG confined in the $c$-axis direction and soft TO$_1$ phonons that are absent in the physical Sr$_2$RuO$_4$; in the next section, we will discuss the candidate material with that has such characteristics but whose 2d band structure projection 
is qualitatively equivalent to that of Sr$_2$RuO$_4$. 


\section{Electron-phonon coupling and superconductivity of the strained osmate-titanate heterostructure}

\begin{figure}
        \includegraphics[width=1.1\columnwidth]{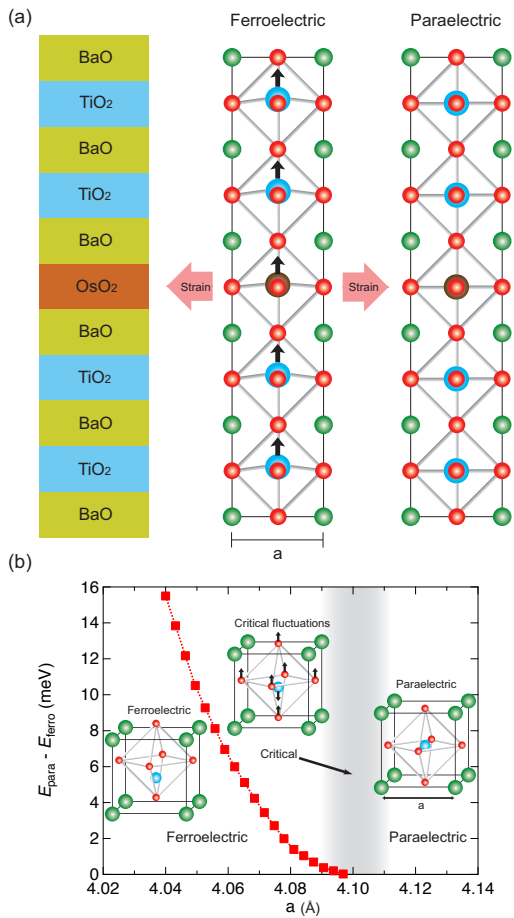}
	\caption{(a) The crystalline structure of the osmate-titanate heterostructure for both the ferroelectric (FE) and the paraelectric (PE) phase; the distance between the osmate layer and the adjacent titanate layer in the PE phase is 4.004\AA. (b) Energy comparison between the PE state with the cubic lattice structure and the (001) FE state shows that the bi-axial tensile strain ($a$ being the $ab$-plane lattice constant) induces change in the crystalline structure of the (001) [BaOsO$_3$][BaTiO$_3$]$_4$ superlattice. For the FE phase, the red square plots the energy saving for the FE ground state compared to the PE state.}
	\label{FIG:lattice}
\end{figure}

We now present the osmate-titanate heterostructure as the physical system that has both the electronic band structure and the electron-phonon coupling qualitatively equivalent the system analyzed in the previous section. 
This heterostructure mainly consists of BaTiO$_3$ with the Os atoms substituting the Ti atoms on a single layer along the $ab$-plane. The single crystal BaTiO$_3$ at low temperature is an electrical insulator that is ferroelectric with the (111) polarization \cite{Dawber2005rev}. Numerical calculation has shown that the $c$-axis component of the BaTiO$_3$ polarization nearly vanishes at zero temperature under the $c$-axis uni-axial compressive strain of 2\% \cite{JHLee2008}, implying that it would be completely suppressed through a continuous quantum phase transition above the critical $ab$-plane bi-axial tensile strain. 
As this transition in the electric polarization is driven by a structural phase transition, we can expect the $c$-axis polarized TO$_1$ phonons to soften 
in its vicinity. Meanwhile, the single osmate layer has been shown to provide the 2DEG with an analogous band structure as Sr$_2$RuO$_4$ \cite{ZZhong2015}, partly due to the Os and Ru atom possessing the same number of valence electron. Hence this heterostructure is a 
fitting candidate to examine for the 2DEG with the band structure analogous to Sr$_2$RuO$_4$ coupled to the soft TO$_1$ phonons; moreover this 2DEG cannot screen the electric field along the $c$-axis that the TO$_1$ phonons would induce.

We shall examine the characteristics of this system through the 
numerical computation of 
the (001) [BaOsO$_3$][BaTiO$_3$]$_4$ superlattice 
shown in Fig.~\ref{FIG:lattice} (a). 
To determine how much tensile strain is required to restore the inversion symmetry with respect to the $ab$-plane, {\it i.e.} turn off the $c$-axis polarization, we calculated the lattice energy 
through the effective lattice Hamiltonian obtained from the local lattice Wannier functions \cite{Rabe1995} for the $c$-axis polarized TO$_1$ phonons. This method not only allows us to determine the relative stability of different electric polarization configurations and the eigenvector (or the relative displacements of different atoms) of the TO$_1$ phonon normal mode but also gives us the various energy parameters for the TO$_1$ phonons. 
Then we calculate the electronic band structure of this superlattice, in both absence and presence of the $c$-axis electric polarization 
from the TO$_1$ normal mode displacement. From this band structure modification, we obtain 
the coupling of electrons to the TO$_1$ phonons 
following the discussion of the last section.

\subsection{Lattice under tensile strain}

\begin{table}[b]
\caption{\label{tab:to1}%
Approximate Os-centered local Wannier function for the TO$_1$ mode and the components of its normalized eigenvectors for the lattice constant of 4.115${\rm \AA}$.
}
\begin{ruledtabular}
\begin{tabular}{cccc}
Element & Site parameters & {\it z} & Eigenvector components \\
\colrule
\multirow{3}{*}{Ba} & 1a & 0 & -0.059 \\ & 2g & 0.2 & -0.023 \\ & 2g & 0.4 & 0.084 \\
\colrule
\multirow{2}{*}{Ti} & 2h & 0.1 & 0.263 \\ & 2h & 0.1 &  0.214 \\ 
\colrule
\multirow{6}{*}{O} & 4i & 0.1 & -0.144 \\ & 4i & 0.3 & -0.144 \\ & 2e & 0.5 &  -0.265 \\ & 1c & 0 & -0.344 \\ & 2h & 0.2 & -0.335 \\ & 2h & 0.4 & -0.228\\
\colrule
Os & 1d & 0.5 & -0.020 \\
\end{tabular}
\end{ruledtabular}
\end{table}

Our calculation of the lattice energy of the (001) [BaOsO$_3$][BaTiO$_3$]$_4$ superlattice indicates that the bi-axial tensile strain 
drives the continuous 
inversion symmetry breaking phase transition. Since the polarization in the $ab$-plane shows no qualitative changes in the vicinity of the critical region, 
it is sufficient for our purpose to compare the lattice energy of the (001) ferroelectric (FE) state and the paraelectric (PE) state with the cubic structure. Given that the (001) FE / PE transition occurs with the softening of the $c$-axis polarized TO$_1$ phonons, we first 
construct, following the lattice Wannier function theory \cite{Rabe1995},  an approximate local basis centered at the Os atom for the $c$-axis polarized TO$_1$ mode as shown in Table~\ref{tab:to1}; 
Fig.~\ref{FIG:band} (b) can be considered as its schematic representation 
with Ru and Sr replaced by Os and Ba, respectively. 
As shown in Fig.~\ref{FIG:lattice} (b), for the  bi-axial tensile strain larger than 2.5\% (or the $ab$-plane lattice constant smaller than 4.10${\rm \AA}$), the effective lattice energy favors the (001) FE state, {\it i.e.} its minima occurs at a finite uniform TO$_1$ mode displacement which vanishes at the 2.5\% tensile strain \footnote{Quantum mechanically, ferroelectricity requires sufficiently small quantum tunneling between the effective lattice energy minima; otherwise we have {\it quantum} paraelectricity.}. On the other hand, any non-zero TO$_1$ mode displacement increases the effecive lattice energy for the bi-axial tensile strain larger than 2.5\% (or the $ab$-plane lattice constant stretched to be larger than 4.10${\rm \AA}$), giving us the PE state. 
Hence we conclude that the TO$_1$ phonons are soft for the $ab$-plane lattice constant around 4.10${\rm \AA}$, and that the tensile strain induces a {\it lattice} quantum criticality 
in the (001) [BaOsO$_3$][BaTiO$_3$]$_4$ superlattice \cite{Chandra2017}; 
this level of strain can be obtained by growing the superlattice on the BaScO$_3$ substrate \cite{Schlom2008}.

Our numerical calculation of the lattice energy allows us to determine the energy parameters of the $c$-axis polarized TO$_1$ phonon modes near the above phase transition to determine the effective lattice Hamiltonian. 
From the local basis shown in Table~\ref{tab:to1}, 
we obtain the effective mass for the TO$_1$ mode, $M_{\rm TO_1}=25.7 {\rm a.m.u.}$, and hence the coefficient for the `kinetic' term of the lattice Hamiltonian. The coefficient of the gradient term 
can be obtained from the phonon velocity $c$, the magnitude of which can be inferred from the BaTiO$_3$ TO$_1$ phonon bandwidth of $\epsilon_c$=37meV \cite{Ghosez1999}. 
This allows us to write 
the effective long-wavelength TO$_1$ phonon Hamiltonian 
\begin{align}
\mathcal{H}_{TO_1} =& \int d^2{\bf r}\left[\frac{1}{2M_{\rm TO_1}}\Pi^2({\bf r}) + \frac{1}{2}M_{\rm TO_1} c^2 \{{\bm \nabla} \tilde{Q}({\bf r})\}^2\right.\nonumber\\ 
&\left. + \frac{1}{2}(\alpha-\alpha_c) \tilde{Q}^2 ({\bf r})+ \frac{1}{4}\beta \tilde{Q}^4({\bf r})  \right],\nonumber
\end{align}
where all lengths are in the unit of the lattice constant, 
$\tilde{Q}$, defined as the $Q$ per unit lattice area, is treated as a field with $\Pi$ being its conjugate momentum, {\it e.g.} $[\Pi({\bf r}), \tilde{Q}({\bf r}')]=-i\hbar \delta^2 ({\bf r}-{\bf r}')$, 
and $\beta>0$ for the quartic coefficient. In terms of this effective Hamiltonian, the quadratic coefficient $\alpha-\alpha_c$ increases with the tensile strain, 
with the polarization on the FE side 
saving the total energy of $L^2 (\alpha_c-\alpha)^2/4\beta$ ($L^2$ being the $ab$-plane area). 
Meanwhile the phonon frequency in the classical harmonic approximation is $\omega_{\bf q}^2 \approx c^2 q^2 + \omega_0^2$ where the phonon frequency gap squared is $\omega_0^2 = (\alpha - \alpha_c)/M_{\rm TO_1}$ for the PE phase and $\omega_0^2 = 2(\alpha_c - \alpha)/M_{\rm TO_1}$ for the FE phase \footnote{From $\mathcal{H}_{TO_1}$, we may consider $\tilde{Q}({\bf q}=0)$ as the ferroelectric order parameter or, equivalently, the paraelectric to ferroelectric transition as the condensation of the TO$_1$ phonon.}; we find 
$\omega_0 = 6.47$ meV at the lattice constant of 4.115\AA. 
We therefore conclude that when the bi-axial tensile strain around the critical value is applied, the TO$_1$ phonon has a very small energy gap and hence dominate the electron-phonon coupling. 

\subsection{Effect of TO$_1$ phonon displacement on the band structure}

\begin{figure}
	\includegraphics[width=1.1\columnwidth]{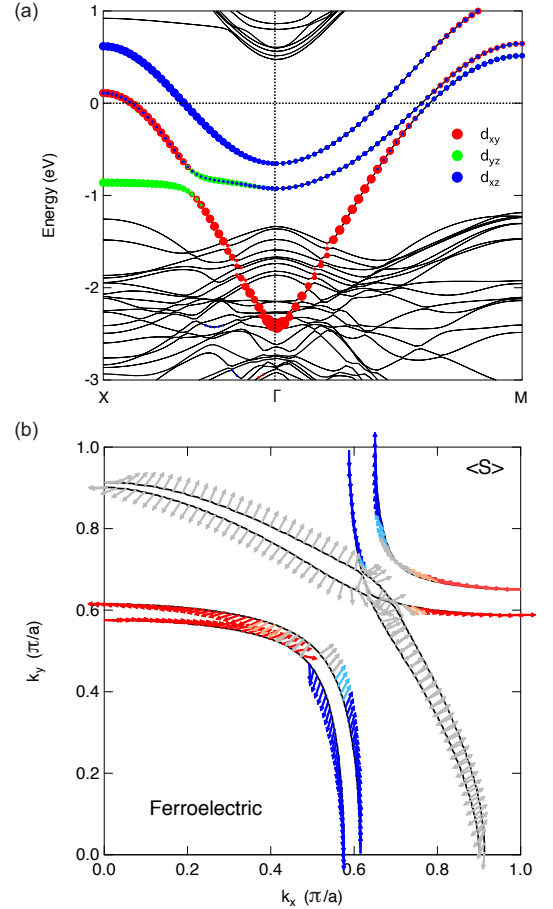}
	\caption{(a) The dispersion and the orbital decomposition of the bands crossing the Fermi level in the (001) [BaOsO$_3$][BaTiO$_3$]$_4$ superlattice without the TO$_1$ phonon mode displacement. (b) The dotted and the solid curves represent the Fermi surfaces of the (001) [BaOsO$_3$][BaTiO$_3$]$_4$ superlattice with the normalized TO$_1$ phonon mode displacement of 0 and 0.25${\rm \AA}$, respectively. 
	The arrows on the Fermi surfaces represent $\langle \hat{\bf S} \rangle$, with the dark blue (red) arrows indicating $|\langle \hat{\bf S} \rangle \cdot \hat{\bf y}|>0.9$ ($|\langle \hat{\bf S} \rangle \cdot \hat{\bf x}|>0.9$) and the light blue (red) arrows $0.8<|\langle \hat{\bf S} \rangle \cdot \hat{\bf y}|<0.9$ ($0.8<|\langle \hat{\bf S} \rangle \cdot \hat{\bf x}|<0.9$).}
	\label{FIG:osmateTO1band}
\end{figure}

We first obtained the band structure of the (001) [BaOsO$_3$][BaTiO$_3$]$_4$ superlattice in the PE state, {\it i.e.} the inversion symmetry preserving case, with the $ab$-plane lattice parameter at 4.115${\rm \AA}$ through the first principle calculation using the density functional theory (DFT) and found the result to be qualitative agreement the Sr$_2$RuO$_4$ model of the previous section. The $k_z$ dispersion and the contribution from the Ti orbitals being both negligible in all cases, we are justified in viewing the (001) [BaOsO$_3$][BaTiO$_3$]$_4$ superlattice as a layered osmate 2DEGs; hence this electronic structure implies the absence of screening for the $c$-axis polarized TO$_1$ phonons. For the PE state, our first-principle calculation, as shown in Fig~\ref{FIG:osmateTO1band} (a), finds three pockets at the Fermi level, 
two of which, originating principally from the $d_{xz/yz}$ orbitals, show quasi-1d dispersion while the remain band, originating principally from the $d_{xy}$ orbital, disperses more isotropically, as in the Sr$_2$RuO$_4$ model. 
As shown in Appendix, there is high overlap between the band basis and the orbital basis over most of the Fermi surface, save for the avoided band crossing regions.

Our DFT calculations of the modified band structure of the (001) [BaOsO$_3$][BaTiO$_3$]$_4$ superlattice with 
finite TO$_1$ mode displacements shows the stronger effects of the inversion symmetry breaking in the quasi-1d bands as in the Sr$_2$RuO$_4$ model of the previous section. The calculations were carried out with the $ab$-plane lattice parameter maintained at 4.115\AA, where the lattice ground state has the inversion symmetry with respect to the $ab$-plane, for the TO$_1$ mode displacement of 0.25\AA, 0.375\AA, and 0.5\AA, respectively (corresponding to the Os atom displacement of 0.005\AA, 0.0075\AA, 0.01\AA\, from the eigenvector components in  Table~\ref{tab:to1}). 
As shown in Appendix, by examining the $k_F$ splitting, we find the magnitude of the Rashba effects on all three bands crossing the Fermi level to be proportional to the TO$_1$ mode displacement. 
The proportionality constants that we obtain between the TO$_1$ mode displacement and the Rashba term of Eq.~\eqref{EQ:hamR_eff} - $\tilde{\lambda}_\alpha=0.233$ eV/\AA, $\tilde{\lambda}_\beta=0.147$~eV/\AA, and $\tilde{\lambda}_\gamma=0.084$~eV/\AA~at the Fermi surface on the $\Gamma$X or the MX segments - 
indicate that the TO$_1$ phonons couples more strongly to the quasi-1d bands. Hence, we conclude that they are the active bands in the TO$_1$ phonon mediated superconductivity.

Furthermore, the Rashba effect produces highly uniform spin textures over most of the quasi-1d Fermi surfaces, an indicator of the weakness of the inter-helicity Cooper pair scattering induced by the TO$_1$ phonons on the quasi-1d bands. 
Fig.~\ref{FIG:osmateTO1band} (b) shows the direction of the 
spin expectation value $\langle \hat{\bf S} \rangle$ of the helicity states on the Fermi surface. 
For most of the Fermi surfaces, we can consider $\langle \hat{\bf S} \rangle$ to be the $\hat{\bf d}^{xz/yz}_{\bf k}$ in Eq.~\eqref{EQ:hamR_eff} given the strong overlap between the band and orbital bases shown in Appendix. 
Therefore, our conclusion from Fig.~\ref{FIG:osmateTO1band} (b) is that the variation of $\hat{\bf d}_{\bf k}$ is weak over most of quasi-1d Fermi surfaces, just as in the Sr$_2$RuO$_4$ model of the previous section. 
We also do not see any strong variation in $|{\bf d}_{\bf k}|$ over the quasi-1d Fermi surfaces, again in agreement with the Sr$_2$RuO$_4$ model. This indicates that the TO$_1$ phonons induce only weak inter-helicity Cooper pair scattering in the quasi-1D bands, and hence, from our discussion in the previous section, the magnitudes of the TO$_1$ phonon induced 
odd parity and $s$-wave pairing interactions are much larger than their difference. 
Hence 
our result suggests that the TRI TSC can be stabilized for the repulsive Hubbard interaction of a wide range of magnitude near the onset of the $c$-axis polarization in the (001) [BaOsO$_3$][BaTiO$_3$]$_4$ superlattice.


Lastly, we find the electronic interaction through the TO$_1$ phonons to be strong enough in the Cooper channel at the $ab$ lattice constant of 4.115${\rm \AA}$ for superconductivity to be accessible, suggesting that the range of the bi-axial strain around the lattice quantum critical point where superconductivity can be observable would not be negligible. This conclusion is based on the standard self-consistency equation for the pairing gap
\begin{align}
1 =& \frac{1}{2N} \sum_{{\bf k}'} V_0 ({\bf k} - {\bf k}') \left(\frac{|{\bf d}_{\bf k}|+|{\bf d}_{{\bf k}'}|}{2}\right)^2\frac{1}{\sqrt{\xi_{k'}^2 + |\Delta|^2}}\nonumber\\ 
=& \frac{1}{2N} \sum_{{\bf k}'} \frac{\tilde{\lambda}^2}{M_{\rm TO_1} \omega_{{\bf k}-{\bf k}'}^2} \sin^2 k_F \frac{1}{\sqrt{\xi_{k'}^2 + |\Delta|^2}},
\end{align}
which is valid for our purpose given the odd-parity / $s$-wave pairing degeneracy in the quasi-1d band with a uniform ${\bf \hat{d}}_{\bf k}$. Using the quasi-1d approximation $ \frac{1}{N} \sum_{{\bf k}'} \to  N_{1d}(0) \int^{+\epsilon_c}_{-\epsilon_c} d\xi_{k'} \int_{-\pi}^\pi dk'_y/2\pi$, where $\xi_k \approx -2t (\cos k - \cos k_F)$ gives us $N_{1d}(0) \approx 1/2\pi t |\sin k_F|$, we obtain the pairing gap magnitude of
\begin{equation}
|\Delta| \approx 2\epsilon_c \exp\left(-\frac{M_{\rm TO_1} \omega_0 \epsilon_c}{\sqrt{2}\hbar |\sin k_F|}\frac{t}{\tilde{\lambda}^2}\right),
\label{EQ:GapEstimate}
\end{equation}
which, when the $\alpha$ band value $\tilde{\lambda}=0.233$eV/\AA, gives us $|\Delta| \approx 500$mK from $\log (2\epsilon_c/|\Delta|) \approx 7.1$, 
suggesting that the superconducting state should be accessible with this lattice constant. 
We note that this is superconductivity outside the lattice quantum criticality in the sense that the FE coherence length $\xi_{FE} = \omega_0/c$ is still only in the order of the lattice constant. Rather, just as in the FE phase Rashba effect \cite{ZZhong2015}, it is the strength of both the lattice-electron coupling and the Os atomic spin-orbit coupling that is giving rise to this robust pairing interaction over an appreciable range of the $ab$-plane lattice constant.

\section{Discussion}

We have shown how an electronic system in vicinity of the inversion symmetry breaking phase transition can be obtained in a superlattice where a 2DEG is sandwiched between insulating layers. This setup aims to maximize the effect of the inversion symmetry breaking, {\it e.g.} the electric polarization, when perpendicular to the 2DEG, would be unscreened thanks to the insulating layers. Hence the insulating layer should be close to the PE / FE transition and induce a strong quantum confinement on the 2DEG \cite{Shanavas2014l}. That perovskite transition metal oxide provides various examples of such insulators, where the transition can be tuned by the bi-axial strain, has been extensively discussed and observed  in recent years \cite{Haeni2004, YJZhang2014}. We note that there were also 2DEG features required specifically for obtaining TRI TSC, {\it e.g.}  an odd number of pockets at the Fermi level in the PE state. However, the 2DEG pocket number is not easily predictable from the constituent atoms. For instance, we find the number of pockets for the (001) [SrRuO$_3$][SrTiO$_3$]$_4$ superlattice is 6 rather than 3, even though the Os and Ru atoms both have the same number of valence electrons (not to mention Ba and Sr). The effective intra-orbital Hubbard interaction is also difficult to determine quantitatively, and any possible instabilities of the 2DEG in which it plays the primary role we have not attempted to address.  
The strain-induced softening of the TO$_1$ phonons coupled to 2DEG may be relevant in other oxide systems such as the LaAlO$_3$ / SrTiO$_3$ heterostructure \cite{Ohtomo2004, Mannhart2010, Takagi2010, Shalom2010}.

This tunable quantum criticality arising from the inversion symmetry breaking coupled to 2DEG promises to be a rich playground for physics of strong correlation \cite{Chandra2017}. In the vicinity of the lattice quantum criticality, the soft TO$_1$ phonon mediated electronic interaction that we are finding raises the possibility of both the high $T_c$ for the superconducting state and the non-Fermi liquid behavior of the normal state. The strong lattice-electron coupling also implies possible shift in the lattice quantum phase transition point, an effect that is neglected in our analysis \cite{Ruhman2017, Shota2018}; we note, however, that these studies have found the lattice phase transition to remain continuous despite the shifts in the transition point. In addition to this strong interaction between electrons in the normal state, there is also possibility of strong interaction between the Bogoliubov-de Gennes quasiparticles in the superconducting state due to 
the existence of a competing pairing channel \cite{Park2015}; these quasiparticles include the helical Majorana edge states for the TRI TSC \cite{ZXLi2017}. As is always the case in the strongly correlated systems, exact numerical methods would be required to quantitatively determine of the TRI TSC stability. 

{\it Acknowledgement}: We would like to thank Daniel Agterberg, Sudip Chakravarty, Gilyoung Cho, Hosub Jin, Changyoung Kim, Heungsik Kim, Minu Kim, Tae Won Noh, Srinivas Raghu, and Hong Yao for sharing their insights. 
M.L., H.-J.L. and J.H.L are supported by Basic Research Laboratory (NRF2017R1A4A1015323) and S.B.C by Basic Science Research
Program through the National Research Foundation of Korea (NRF) funded by the Ministry of Education (2018R1D1A1B07045899 and 2018R1A6A1A06024977).

\appendix

\section{Fermi surface parameters}

\begin{figure}
	\includegraphics[width=0.9\columnwidth]{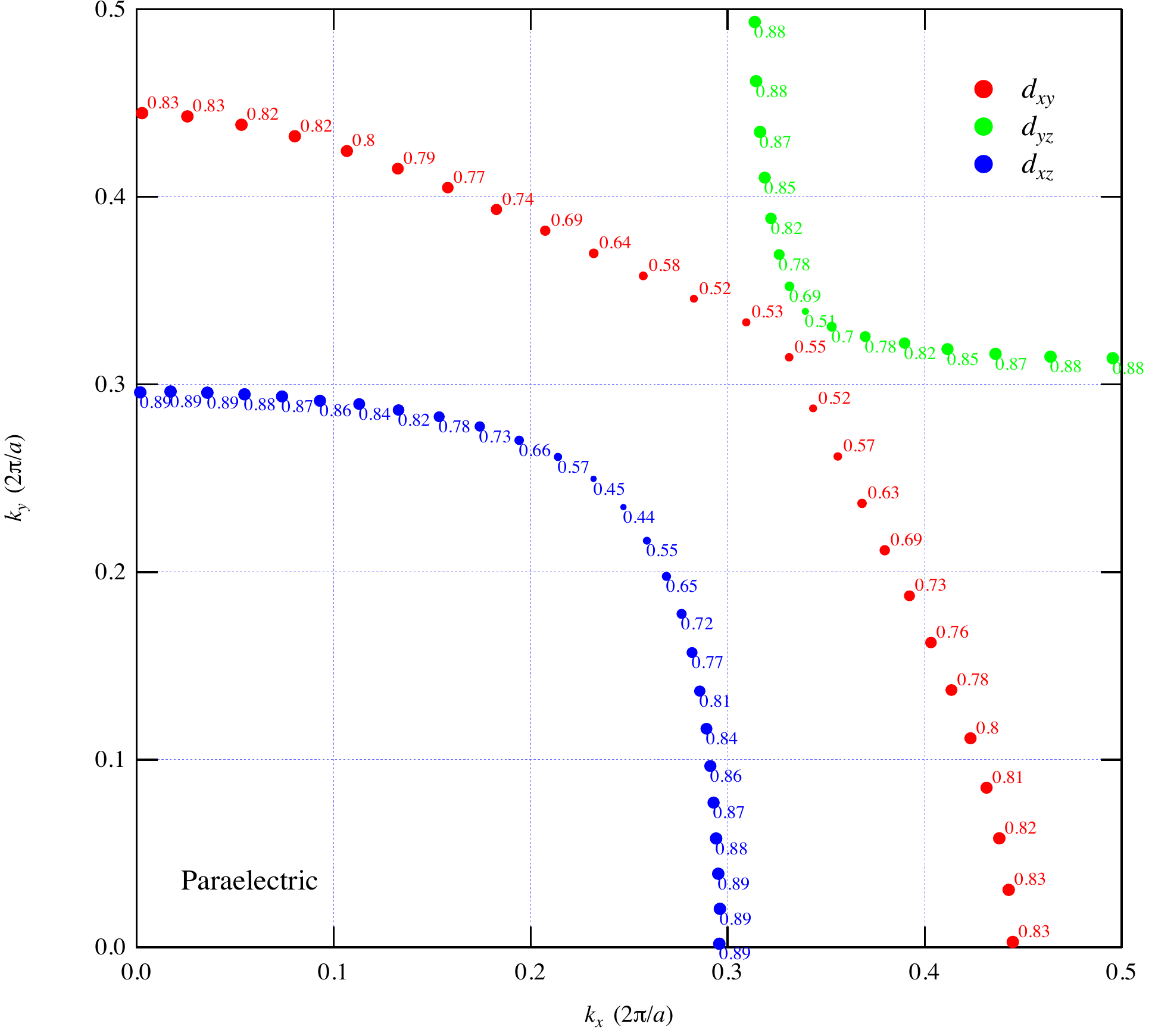}
	\caption{This plot shows by color which $t_{2g}$ orbitals make the dominant contribution to the eigenstates at the Fermi level for the paraelectric state. The number shows the percentage of this orbital's contribution out of the total overlap with the $t_{2g}$ orbitals.}
	\label{FIG:FSOrbital}
\end{figure}

Fig.~\ref{FIG:FSOrbital} shows that the eigenstates on the Fermi surface largely overlaps with the orbital eigenstate in the paraelectric state; this justifies using identifying ${\bf \hat{d}}_{\bf k}$ with $\langle {\bf \hat{S}} \rangle$ in our analysis.

\begin{table}[bt]
\caption{$k_F$ splitting from the varying TO$_1$ mode displacement for the lattice constant of 4.115\AA.}
\begin{ruledtabular}
\begin{tabular}{cccc}
Band & Q=0.25\AA & Q=0.375\AA & Q=0.5\AA \\
\colrule
\multirow{2}{*}{$\alpha$} & 1.8014 & 1.9151 & 1.9396 \\  & 1.6336 & 1.6355 & 1.5815 \\ 
\colrule
\multirow{2}{*}{$\beta$} & 1.6022 & 1.5601 & 1.5180 \\ & 1.7078 & 1.7291 &  1.7499 \\ 
\colrule
\multirow{2}{*}{$\gamma$} & 2.5937 & 2.5937 & 2.6144 \\ & 2.6565 & 2.6986 & 2.7407 \\ 
\end{tabular}
\end{ruledtabular}
\end{table}

The splitting of the Fermi surface allows us to determine the Rashba coefficient. Using 
the Rashba term of Eq.~\eqref{EQ:hamR_eff} 
and approximating the band structure of the paraelectric state arises solely from the nearest-neighbor hopping $t$ of Eq.~\eqref{EQ:invSymm}, we obtain $\lambda=t\Delta k_F$; for the (001) [BaRuO$_3$][BaTiO$_3$]$_4$ superlattice, the tight-binding fitting gives us $t=0.3474$eV. The $k_F$'s for various non-zero value of the TO$_1$ mode displacement shown in Table II is for the $\alpha$ band on the MX segment and the $\beta$, $\gamma$ bands on the $\Gamma$X segment.

\bibliography{TO1}

\end{document}